# Climate Dynamics: A Network-Based Approach for the Analysis of Global Precipitation

**Stefania Scarsoglio[1]\*, Francesco Laio[2], Luca Ridolfi[2]**

**1** Department of Mechanical and Aerospace Engineering, Politecnico di Torino, Torino, Italy, **2** Department of Environment, Land and Infrastructure Engineering, Politecnico di Torino, Torino, Italy

## Abstract

Precipitation is one of the most important meteorological variables for defining the climate dynamics, but the spatial patterns of precipitation have not been fully investigated yet. The complex network theory, which provides a robust tool to investigate the statistical interdependence of many interacting elements, is used here to analyze the spatial dynamics of annual precipitation over seventy years (1941–2010). The precipitation network is built associating a node to a geographical region, which has a temporal distribution of precipitation, and identifying possible links among nodes through the correlation function. The precipitation network reveals significant spatial variability with barely connected regions, as Eastern China and Japan, and highly connected regions, such as the African Sahel, Eastern Australia and, to a lesser extent, Northern Europe. Sahel and Eastern Australia are remarkably dry regions, where low amounts of rainfall are uniformly distributed on continental scales and small-scale extreme events are rare. As a consequence, the precipitation gradient is low, making these regions well connected on a large spatial scale. On the contrary, the Asiatic South-East is often reached by extreme events such as monsoons, tropical cyclones and heat waves, which can all contribute to reduce the correlation to the short-range scale only. Some patterns emerging between mid-latitude and tropical regions suggest a possible impact of the propagation of planetary waves on precipitation at a global scale. Other links can be qualitatively associated to the atmospheric and oceanic circulation. To analyze the sensitivity of the network to the physical closeness of the nodes, short-term connections are broken. The African Sahel, Eastern Australia and Northern Europe regions again appear as the supernodes of the network, confirming furthermore their long-range connection structure. Almost all North-American and Asian nodes vanish, revealing that extreme events can enhance high precipitation gradients, leading to a systematic absence of long-range patterns.

**Funding:** Funding from the Italian Ministry of Education, Universities and Research (Futuro in Ricerca 2012, Project RBFR12BA3Y) is acknowledged. The funders had no role in study design, data collection and analysis, decision to publish, or preparation of the manuscript.

**Competing Interests:** The authors have declared that no competing interests exist.

\* E-mail: stefania.scarsoglio@polito.it

## Introduction

By combining graph theory and statistical physics, complex network theory provides a powerful tool to investigate the structure and function of complex systems with a large number of interacting elements. The development and characterization of complex networks [1–3] makes their application suitable to analyze a wide range of systems from nature to economy, from engineering to society [4]. Beside the well-established applications to Internet and World Wide Web, neural connections and social dynamics [5], complex networks have been successfully used to study many different phenomena such as, for example, human migration [6], cancer metastasis [7] and earthquake occurrence [8].

The extension of complex network theory to climate sciences is a very recent area yielding *climate networks*, which usually rely on gridded time series of meteorological preprocessed variables. The nodes of the network are identified by geographical regions corresponding to single points of measurement on the spatial grid of the underlying climate database. Each node has a measured state variable which varies in time. A link between two nodes exists if there is a significant statistical interdependence between their time series. The linear cross-correlation function is typically used as the simplest possible measure of the statistical interdependence of the temporal series. However, the influence of the choice of an association measure on the topology of the climate network has been studied, by accounting how the temporal complexity of time series influences the absolute correlations [9] and proposing alternative criteria based on the nonlinear mutual information [10,11].

Until now, attention has been mainly devoted to networks based on the global surface temperature field to understand the influence of El Niño and La Niña events in regions which are far from the El Niño-Southern Oscillation (ENSO) area. Although the temperatures in different zones of the world are not significantly affected by El Niño and La Niña, it was surprisingly found that the climate network during these events is sensitively influenced by showing a different structure and a consistent amount of broken links [12–14]. The community structure [15] as well as the dynamics of interacting networks [16] have been investigated using the surface temperature fields and related variables (e.g., sea level pressure and equipotential heights).

Apart from some works where different meteorological variables such as equipotential heights [15–17], sea surface temperature,

humidity, and related data [18] are also analyzed, the great part of climate network literature deals with surface air temperature data only. In particular, to the best of our knowledge, a global precipitation analysis has not yet been performed by using the complex network theory. Only Malik *et al.* [19] recently carried out a complex network study of local extreme monsoonal rainfall in South Asia, while Bayesian networks have been employed to analyze local precipitation in the Iberian peninsula [20,21].

The gap concerning global precipitation in climate networks is evident even though precipitation teleconnections, especially related to the ENSO occurrence [22,23], have been recognized for being a crucial aspect in climate and hydrological changes [24], and are increasingly examined for Asian monsoons [25–28], European [29,30], African [31] and American [32,33] rainfall events.

The present work arises in this scenario and aims at being a first step towards filling this gap in climate network analysis; in fact, precipitation, together with surface temperature and wind, atmospheric pressure and humidity, is one of the most important meteorological variables in defining the climate dynamics. The global annual precipitation over seventy years (1941–2010) is analyzed by means of the complex network theory. We use the Global Precipitation Climatology Centre (GPCC) Database [34–36], which is one of the most reliable precipitation datasets providing land-surface precipitation from rain-gauges over the period 1901–2010. Pre-processing of the data is performed to (i) define nodes corresponding to geographical regions (cells) with the same area, and (ii) consider data mainly based on *in situ* observations (rather than on interpolated values). The precipitation network is built identifying a node with a geographical region, which has a temporal distribution of measured precipitation, and using the linear correlation function to evaluate possible links between nodes. If the statistical interdependence between two nodes is above a suitably chosen threshold, a link between the two nodes is established. The precipitation network is described through classical tools of the complex network theory - such as the degree centrality, the betweenness centrality and the clustering coefficient - as well as measures introduced here for the first time: the weighted average topological distance, which generalizes the average topological distance definition, and the average physical distance of a node from the rest of the network. To analyze the sensitivity of the network to the physical closeness of the nodes, short-term connections are broken and edges between physically distant nodes only survive. In so doing, the unavoidable spatial correlation between physical neighbors is left aside in favor of highlighting the possible interdependence between not confining regions.

## Materials and Methods

In this section the database used to define the precipitation network is described. Details on the pre-processing analysis of data are then offered. Afterwards, we summarize some fundamental concepts in complex network theory [3,5]. We only introduce measures which are relevant to the present analysis. In the end, starting from the spatio-temporal global precipitation distribution, details on how to build the precipitation network will be given.

### GPCC Full Database Description

The present investigation uses the Global Precipitation Climatology Centre (GPCC) Full Data Reanalysis Version 6, which consists of monthly land-surface precipitation data from rain-gauges built on Global Telecommunication System (GTS) and historic data [34–36]. The GPCC Full Database covers the period from January 1901 to December 2010 and is based on both real-time raingauge data as well as non real-time sets of data. The new extended database version was released in December 2011 and the data coverage per month varies from 10,700 to more than 47,000 stations. Non real-time data, coming from dense national observation networks of individual countries and other global and regional collections of climate data, are integrated in the GPCC Full Database using the GPCC Precipitation Climatology Database as analysis background (for more details, see the online documentation [36]).

The monthly global precipitation is reported on a regular grid with a spatial resolution of $0.5°\times0.5°$, $1.0°\times1.0°$, and $2.5°\times2.5°$ latitude by longitude. The global gridsystem goes from $(-180°$ W, $+90°$ N) to $(+180°$ E, $-90°$ S). For every gridcell three kinds of data are given: (i) monthly precipitation [mm/month]; (ii) mean monthly precipitation [mm/month] based on the GPCC Precipitation Climatology; (iii) number of gauges per grid. We obtain the annual precipitation by summing the monthly precipitation values.

The GPCC Full Database provides one of the most accurate and complete *in situ* precipitation data sets. In fact, the wide spatial and temporal coverage makes this database suitable for verification of models [37,38], for analysis of historic global precipitation [39–43], and for research concerning the global water cycle [44,45], e.g., trend and time-series analyses [46,47]. Moreover, the number of stations per gridcell is an additional useful piece of information, which can be exploited when evaluating the spatio-temporal precipitation distribution.

### Data Pre-Processing

Although the GPCC Full Database is a very accurate and detailed dataset, there are two main concerns to use it, as it is, to define a complex network. In this section we discuss these two issues and propose our solutions to overcome them.

First, the regular grid based on the angular partition of the terrestrial surface (i.e., the geographic coordinate system) leads to the definition of gridcells with different geometric area. This heterogeneity, which becomes more evident by approaching regions far from the equator, may induce substantial bias and spurious correlation when building the precipitation network. One way to avoid this bias is to use a tessellation technique [48] to divide the gridcells into suitable two-dimensional structures. An alternative axiomatic scheme, based on the idea of node splitting invariance to obtain consistent weights for the most commonly used network parameters, was proposed by Heitzig *et al.* [49].

Here, we adopt a simpler approach: we build a new grid system with all square cells having a fixed area, $d_e \times d_e$. We focus on the equator, where the original GPCC grid system with $2.5°\times2.5°$ gridcell resolution yields a square cell dimension $(d_e \times d_e)$ of $278.3\times278.3$ km$^2$. The new graticule, which is here proposed, is built maintaining this cell dimension $(d_e \times d_e)$ fixed for all latitudes. As a consequence, the number of the new cells is variable over the latitude and, in particular, the new graticule has fewer cells than the original GPCC grid system when moving far from the equator. The number of raingauges of a new cell is the sum of all the raingauges present in the GPCC cells which are completely contained by the new cell. The precipitation value of the new cell is instead the average of the precipitation values measured by those GPCC cells which are entirely included into the new cell. To avoid possible overlaps in the longitudinal direction between cells of the new graticule, we adopt the following convention: when two new cells share an edge falling into an original GPCC cell, the contribute of the original cell (in terms of number of raingauges and precipitation value) is fully allocated to the cell of the new grid system which covers most of the GPCC cell area.

The second concern is about the spatio-temporal distribution of measurement stations. In the GPCC Full Database, in fact, there exists an amount of cells for which no measurement is available over several months (i.e., the number of raingauges for the gridcell is 0). However, in these cases, the global precipitation information are recovered through the interpolation of global and mean data offered by the GPCC Precipitation Climatology Database. In so doing, the spatio-temporal coverage is complete but some precipitation values can be fully based on interpolated data. This aspect becomes important when the oldest data are analyzed, since fewer measurements were available. In order to consider data mainly based on in situ observations, we define a grid cell as active for a fixed year if there is at least one measurement for every month of the year. Otherwise the grid cell is not active. We then restrict the analysis to a temporal window of 70 years starting from 1941 to 2010, and consider only cells which are active (also not consecutively) for at least 50 years over the temporal window of 70 years. In this way, more than 90% of the stations will be included in the spatio-temporal precipitation distribution and, in the worst cases, less than 30% of the distribution (20 years over 70) will rely on interpolated data. The number of active cells is $M = 1731$.

## Complex Network Tools: Definitions and Structural Properties

A network (or graph) is defined by a set $V = 1,...,N$ of nodes and a set $E$ of edges (or links) $\{i,j\}$. Here, we assume $N \leq M$, that is the number of nodes of the network, $N$, can be equal or lower than the number of active cells, $M$. Moreover, we suppose that only one edge can exist between a pair of nodes and no self-loops $\{i,i\}$ is allowed. The *adjacency matrix*, $A$:

$$A_{ij} = \begin{cases} 0, & \text{if } \{i,j\} \notin E \\ 1, & \text{if } \{i,j\} \in E, \end{cases} \quad (1)$$

takes into account whether a link is active or not between nodes $i$ and $j$. Since the network is considered as undirected, $A$ is symmetric. Since no self-loops are allowed, $A_{ii} = 0$.

The *degree centrality* of a node is defined as

$$k_i = \frac{\sum_{j=1}^{N} A_{ij}}{N-1}, \quad (2)$$

and gives the number of first neighbors of the node $i$, normalized over the total number of possible neighbors $(N-1)$. The *degree distribution*, $p(k)$, defines the fraction of nodes in the graph having degree $k$. In other words, the degree distribution is the probability that a node in the network is connected to $k$ other nodes.

We here propose the *weighted average topological distance* of a node as

$$\bar{D}_i = \frac{\sum_{j \in nn(i)} d_{ij}}{N_{ci}} \frac{N-1}{N_{ci}}, \quad (3)$$

where the shortest path length, $d_{ij}$, is the minimum number of edges that have to be crossed from node $i$ to node $j$, and $nn(i)$ is the set of all neighbors of $i$. $N_{ci}$ is the number of nodes connected to node $i$ ($N_{ci} \in [1, N-1]$). The first ratio of the right hand side of Eq. (3) accounts for the mean topological distance of node $i$ with respect to all the $N_{ci}$ nodes linked to it, while the second ratio is a weight coefficient considering how strongly node $i$ is connected to the rest of the graph. We introduce this notation to generalize the classical average topological distance definition [50], $\bar{d}_i = \sum_{j=1}^{N} d_{ij}/(N-1)$. In fact, when the graph has disconnected components the average topological distance definition diverges. Relation (3) is identical to the average topological distance when the graph is completely connected ($N_{ci} = N-1$ for every node). In the case of a graph with disconnected nodes, instead, $\bar{D}_i$ does not diverge and can vary in the interval $[1, N-1]$. The extreme value $N-1$ is reached when two nodes, $i$ and $j$, are directly connected one to each other ($d_{ij} = 1$), but disconnected from the other nodes ($N_{ci} = 1$). A large $\bar{D}_i$ value means that node $i$ is topologically far from the rest of the network.

The *local clustering coefficient* of a node is

$$C_i = \frac{e(\Gamma_i)}{\frac{k_i(k_i-1)}{2}}, \quad (4)$$

where $\Gamma_i$ is the set of first neighbors of $i$, $e(\Gamma_i)$ is the number of edges connecting the vertices within the neighborhood $\Gamma_i$, and $k_i(k_i-1)/2$ is the maximum number of edges in $\Gamma_i$, $0 \leq C_i \leq 1$. The local clustering coefficient gives the probability that two randomly chosen neighbors of $i$ are also neighbors. The *global clustering coefficient* is the mean value of $C_i$, $\bar{C} = \sum_{i=1}^{N} C_i/N$.

The *betweenness centrality* of a node is

$$BC_k = \sum_{i,j \neq k} \frac{\sigma_{ij}(k)}{\sigma_{ij}}, \quad (5)$$

where $\sigma_{ij}$ are the number of shortest paths connecting nodes $i$ and $j$, while $\sigma_{ij}(k)$ gives the number of shortest paths from $i$ to $j$ crossing node $k$. If node $k$ is traversed by a large number of all existing shortest paths (that is, if $BC_k$ is large), then node $k$ can be considered an important mediator for the information transport in the network.

## Building the precipitation network

The number of active cells, as previously described, is $M = 1731$. Once the time series of the annual precipitation is obtained for each active cell, we can evaluate the cross correlation between all pairs of them. We use the linear Pearson correlation as it is the simplest possible measure to quantify the degree of statistical interdependence between the temporal series. Moreover, Donges *et al.* [10] found a high level of similarity between Pearson correlation and mutual information networks. The correlation coefficient is given by an element of the correlation matrix, $R_{ij}$, which is symmetric and estimates the strength of a linear interdependence between two temporal series, $s_i$ and $s_j$ ($i,j \in [1,M]$, $R_{ii} = 1$ by definition). The correlation coefficient can vary between $-1$ and $1$. A large positive value means the temporal series are strongly correlated, while a large negative value indicates a strong anti-correlation. Since we are interested in both large positive and negative correlation values, the absolute value of $R_{ij}$ will be used to build the precipitation network [10]. Moreover, the physical distance, $l_{ij}$, is evaluated in kilometers as the shorter great circle path between nodes $i$ and $j$ and stored in the symmetric matrix $L_{ij}$.

The *average physical distance* of an active cell (node) is defined here as

$$\bar{L}_i = \frac{\sum_{j=1}^{M} l_{ij}}{M-1}, \quad (6)$$

where $M$ is the number of active cells and $l_{ij}$ is the physical distance between nodes $i$ and $j$ defined above ($l_{ii} = 0$ by definition).

The *edge density* $\rho(\tau)$ is defined as:

$$\rho(\tau) = 1 - P_R(\tau) = \frac{n(\tau)}{\frac{N(N-1)}{2}}, \qquad (7)$$

where $n(\tau)$ is the number of active links (edges) when the absolute value of $R$ (in the following we abbreviate the correlation $R_{ij}$ with $R$) is above the threshold $\tau$, while $P_R(\tau)$ is the cumulative distribution function of the correlation $R$.

To define the adjacency matrix, $A$, and therefore the network, we refer to Eq. (1) and define that an edge, $E$, between nodes $i$ and $j$ exists when $|R_{ij}| > \tau$. In so doing, the resulting precipitation network is undirected (A is symmetric) and unweighted (all the $|R_{ij}|$ values above the threshold correspond to $A_{ij} = 1$). The selection of the threshold $\tau$ is a non-trivial aspect of building a climate network [10,12,13]. Since we are primarily interested in highlighting strong correlated and anti-correlated connections, we set $\tau = 0.5$. With this threshold value, the number of active links is $n = 9481$ and the number of nodes is $N = 1674$ (57 nodes out of 1731 are not connected with any other node of the network). The chosen threshold, $\tau = 0.5$, corresponds to an edge density value, $\rho$, equal to 0.0068. We graphically make the nodes coincide with the center of each active cell, see the blue symbols in the left panel of Fig. 1.

## Results

The properties of the precipitation network are here presented and discussed. Among the network measures introduced in the *Materials and Methods* Section, particular attention will be paid to the degree centrality and the weighted average topological distance. In fact, these two parameters reveal to be the most meaningful for the present analysis.

As mentioned, the precipitation network is made of $N = 1674$ nodes. However, not all of them are completely connected one to the other. This means that the graph has disconnected components. This aspect justifies the choice of proposing a different definition of the mean topological distance, see Eq. (3). In particular, 1632 nodes are completely linked and form a big subnetwork, while the remaining 42 nodes contribute to create 14 smaller micro-networks. The size (number of nodes) of each micro-network varies from 9 to 2. Although visible through the betweenness centrality, the best parameter which physically individuates the nodes of these smaller networks is the weighted average topological distance, $\bar{D}_i$.

The data analysis is completed by describing the properties of another precipitation network with an additional constraint: an edge, $E$, between nodes $i$ and $j$ exists when $|R_{ij}| > \tau$ and $l_{ij} > \delta$. A suitably large value of the threshold $\delta$ leads to define a new precipitation network, where edges only exist between nodes which are physically far from each other. In so doing, the unavoidable spatial correlation between physical neighbors is left aside in favor of highlighting the possible interdependence between not confining regions (i.e., $\delta > d_e$).

To focus on possible links between regions physically far from each other, we set $\delta = 1000$ km and obtain a new network with fewer nodes, $N_\delta = 640$, and links, $n_\delta = 1632$, than the network previously discussed. The American and Asiatic continents are the regions which much suffer of the reduction of nodes, while Western Europe and Australia still have an appreciable number of nodes (see the red symbols in the left panel of Fig. 1). The number of active links, $n$, is reported in the right panel of Fig. 1 as a function of the physical distance, $l$ (the scale of values is in $10^3$ km, the red line represents the threshold $\delta = 1000$ km).

## Properties of the precipitation network

We start considering the degree centrality for the basic network, see Fig. 2. One clearly distinguishes two regions with the highest degree centrality values: the Sahel region in Africa, and Eastern Australia. The nodes in these regions are directly connected to a great number of other nodes of the network, therefore they are usually referred to as *supernodes*. Beside the supernode areas, there exist regions with fairly high degree centrality values: Northern Europe, Central Asia, Southern Africa, Western US, and Northeastern Brazil. It should be noted the great difference in terms of degree centrality values, $k_i$, when moving from the West to the East Coast of the US, as well as from Northern Europe towards the Mediterranean Sea. We recall that the degree centrality is the ratio between the number of cells directly linked to a fixed cell, normalized over the total number of possible neighboring cells ($N-1$). Speaking in terms of the physical area directly connected to a cell, the maximum degree centrality value here reached, $k_i = 0.033$, corresponds to a directly connected physical area of about $4.3 \cdot 10^6$ km$^2$, equivalent to the total area of the countries in the European Union.

Real networks are often scale-free, that is power-law degree distributions are displayed [5], with exponents ranging between $-2$ and $-3$. These networks usually result in the simultaneous presence of few nodes highly connected to the others (i.e., *supernodes*) and a great amount of barely connected cells. However, because of the finite size of the network, data can have a rather strong intrinsic noise. To smooth the fluctuations generally present in the tails of the distribution, it is often verified if the cumulative distribution function, $P_K(k)$, presents a power-law behavior. Figure 3 reports the exceedance probability of the node degree, $1 - P_K(k)$. A power law decay with exponent equal to $-2$ is clearly observable in the intermediate range, $k \in [0.005, 0.025]$ (see the red line in Fig. 3).

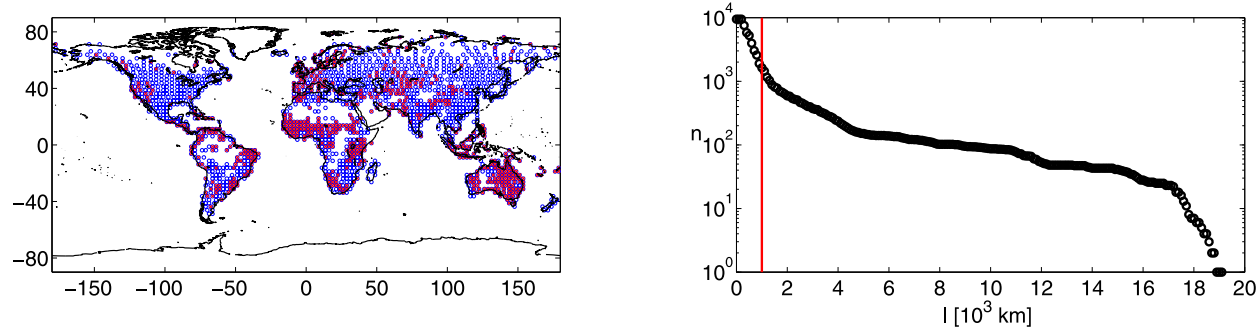

**Figure 1. Nodes and links of the network.** (left) Nodes of the network. Each node graphically coincides with the center of an active spatial cell. Red symbols correspond to the nodes of the network ($N_\delta = 640$) with the additional physical constraint, $l_{ij} > \delta = 1000$ km, while blue symbols correspond to the nodes of the network ($N = 1674$) without this constraint. (right) Number of links, $n$, as a function of the physical distance, $l$ [$10^3$ km]. The scale of values is in $10^3$ km, the red line indicates the physical threshold, $\delta = 1000$ km.
doi:10.1371/journal.pone.0071129.g001

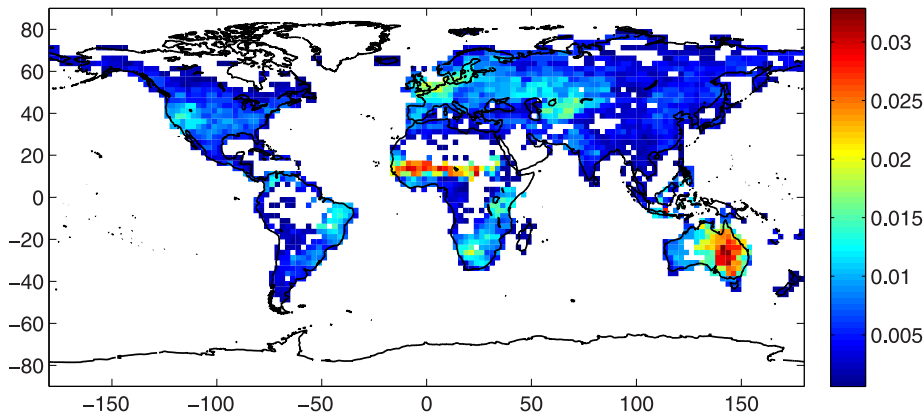

**Figure 2. Degree centrality, $k_i$.** Large values correspond to highly connected nodes.
doi:10.1371/journal.pone.0071129.g002

The weighted average topological distance, $\bar{D}_i$, is represented in the top panel of Fig. 4. The scale of values is restricted to the interval $[6,12]$, while higher values are reported without distinction with the grey color. The grey-colored regions individuate the nodes of the micro-networks. In fact, as mentioned in the *Materials and Methods* Section, disconnected components of the graph have extremely high $\bar{D}_i$ values. In the worst case, when a micro-network has two nodes only, $\bar{D}_i = N - 1 = 1673$. This situation occurs for 18 nodes out of 1674.

As a first comment, there is quite a notable correspondence between high degree centrality and low weighted average topological distance, and viceversa. This is especially evident, on one hand, for large $k_i$ values as in the supernode areas, whose nodes have the lowest $\bar{D}_i$ values. On the other hand, regions with the highest $\bar{D}_i$ values (grey and red colored zones in the top panel of Fig. 4) have $k_i \to 0$. Few remarkable exceptions to this inverse correspondence are the Atlantic Coast of South-America, the

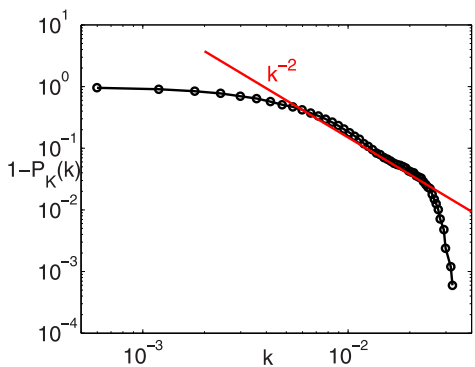

**Figure 3. Exceedance probability of the node degree, $1 - P_K(k)$.**
doi:10.1371/journal.pone.0071129.g003

Mongolian area and the Indonesian archipelago. For these regions medium-low $\bar{D}_i$ values do not correspond to appreciably high degree centrality values. It should be noted that the North-American and European regions have quite the same low $\bar{D}_i$ values.

At this stage, it is useful to compare the weighted average topological distance map, $\bar{D}_i$, with the average physical distance map, $\bar{L}_i$, offered in the bottom panel of Fig. 4 (the scale of values is in $10^3$ km). Some regions (Northwestern Europe, Central Asia and Mongolian area, African Sahel region) have both measures with quite low values and one can infer that the strong topological connection is partially due to the high physical closeness of the nodes involved. However, leaving aside these regions, for the rest of the network there exists an inverse correspondence between the weighted average topological distance and the average physical distance. This aspect is more marked in the Southern Hemisphere, where the average physical distance between active cells is in general higher and, at the same time, nodes are often topologically close one to each other. To this end, one can refer in particular to the Atlantic Coast of South America and Eastern Australia, but also South-Africa, Mid-Western US, Northern South-America and the Indonesian Archipelago. Nevertheless, the inverse proportionality between $\bar{D}_i$ and $\bar{L}_i$ is visible in topologically low connected areas such as Eastern Asia, which is instead a region whose cells on average are not physically far from the rest of the network.

To carry out a sensitivity analysis of the network with respect to the physical neighborhood of the nodes, we here define that an edge between nodes $i$ and $j$ exists if $|R_{ij}| > \tau$ and, at the same time, the additional physical constraint, $l_{ij} > \delta = 1000$ km, is satisfied. In so doing, a different network is specified where edges between nodes distant less than $\delta$ cannot exist.

The degree centrality and the weighted average topological distance are presented in the left and right panels of Fig. 5, respectively. The current scenario deeply emphasizes the role of the supernode areas of the original network (the African Sahel region and Eastern Australia), which are still the regions with the highest degree centrality and the lowest weighted average topological distance. All the other previously existing links are instead weakened or, in several cases, even broken, meaning that their correlation was due to the physical closeness of the nodes involved. In particular, it is worth noting that the US and Western Europe show now very different patterns. Indeed, Western Europe still preserves a large number of highly connected nodes, while the US have a small amount of nodes which are scarcely connected to the rest of the network.

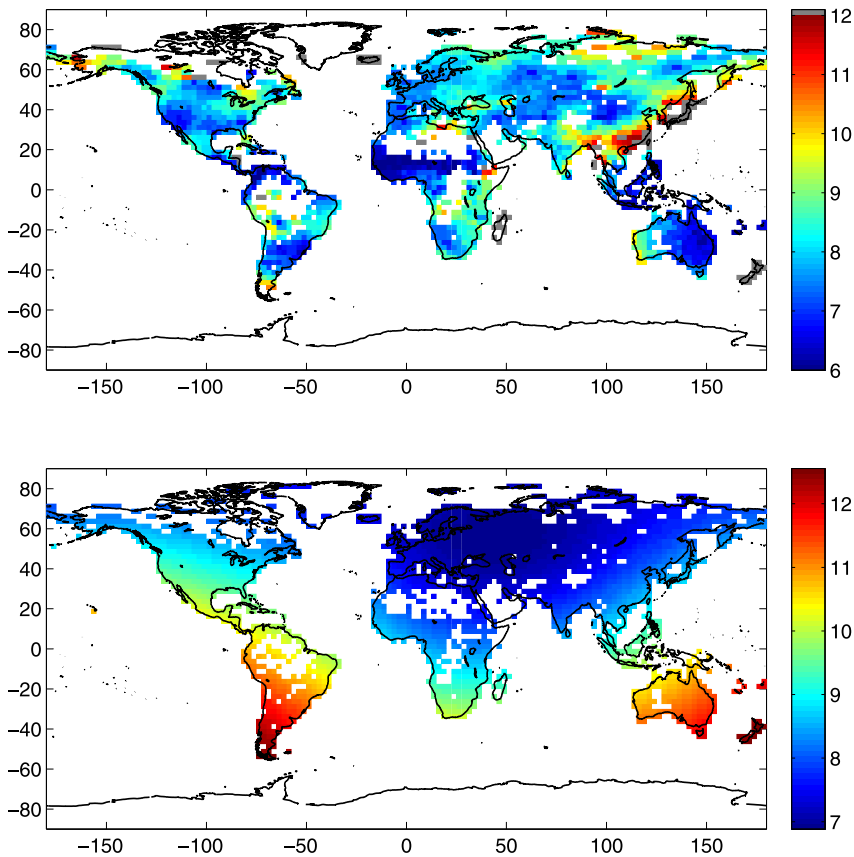

**Figure 4. Weighted average topological distance and average physical distance.** (top) Weighted average topological distance, $\bar{D}_i$. The scale of values spans in the interval [6,12]. Higher values are reported without distinction with grey color. (bottom) Average physical distance, $\bar{L}_i$. The scale of values is in $10^3$ km.
doi:10.1371/journal.pone.0071129.g004

The evident absence of nodes in Asia and North-America (and the poor connection of the few remaining cells) can be thought in terms of the presence of strong precipitation variation on a relatively short spatial scale, thereby leading to the emergence of high precipitation gradients. A high precipitation gradient can be, for instance, enhanced by the occurrence of regional extreme events (e.g., tropical cyclones, monsoons, tornadoes, blizzards, heat waves) which are usually localized in time and space. The Asian and North-American continents, due to their huge land mass extension, experience the most imponent extreme phenomena [51]. Therefore, in these places precipitation strongly varies on regions which are relatively close one to the other, allowing only short-range links to survive, which are eventually broken by the additional physical constraint, $l_{ij} > 1000$ km. Examples of these high precipitation gradient areas are South-East and North-West China, Central Asia (including Mongolia, Kazakhstan and Central Russia), India, Nepal and Pakistan, as well as Western (Washington, Oregon and California), Central (Texas, Louisiana, Oklahoma, Arkansas, Kansas, Nebraska, Missouri, Iowa and Minnesota), and Southeastern (Florida, Mississippi, Alabama) United

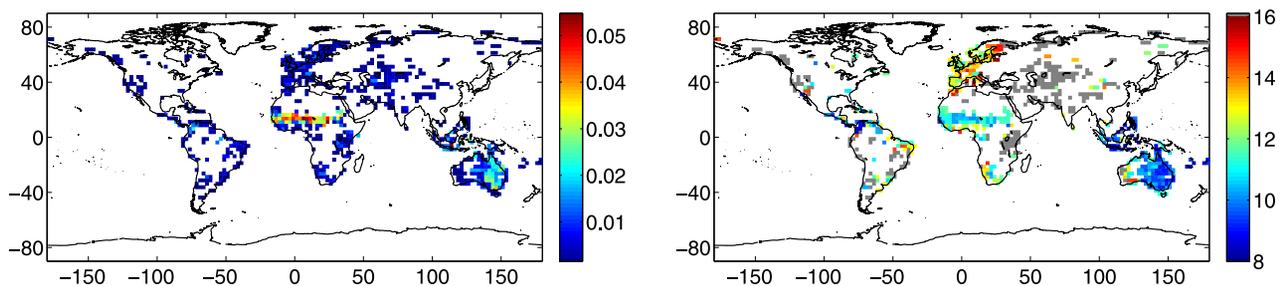

**Figure 5. Long-range network, $l_{ij} > d = 1000$ km.** (left) Degree centrality. (right) Weighted average topological distance. The scale of values spans in the interval [8,16], while higher values are reported without distinction with grey color.
doi:10.1371/journal.pone.0071129.g005

States. In all these cases, the high precipitation gradient makes extremely dry and extremely wet regions coexist in a few hundred kilometers range.

In the supernode regions (Sahel, Eastern Australia and Western Europe) extreme events in general occur less frequently. Medium (Western Europe) or very low (Sahel, Eastern Australia) rainfall spread out more uniformly on a continental scale, the precipitation gradient is weaker and nodes remain connected in the long-range.

Coming back to the original network with $N = 1674$ nodes, the local clustering coefficient and the betweenness centrality are presented in the top and bottom panels of Fig. 6, respectively (note that a logarithmic representation is adopted for the betweenness centrality). These two measures are a little less significant and weakly related to the degree centrality and the weighted average topological distance maps. In fact, both distributions in Fig. 6 are spotted worldwide without evidencing regions of particular interest.

From a qualitative point of view, the local clustering coefficient presents a strong heterogeneity in the central part of Africa, in Eastern Asia and South America. Some patterned zones with high $C_i$ values are found on coastal regions: Brazil, Eastern Australia and Eastern Africa. More in general, $C_i$ seems to mainly vary around the values represented by the green and yellow tones ($C_i \in [0.45, 0.65]$). As plotted in the middle panel of Fig. 6, the probability density function of the local clustering coefficient, $p(C)$, quantitatively confirms this behavior by showing a moderately trimodal distribution. The central mode, by far the most probable one, lies very close to the global clustering coefficient, $\bar{C} = 0.5233$ (see the red line in the middle panel of Fig. 6), which is the arithmetic mean of the $C_i$ values. Going back to the meaning of this parameter, the present results can be interpreted as follows: on average, there is about 52% of chances that two randomly chosen neighbors of node $i$ are also neighbors.

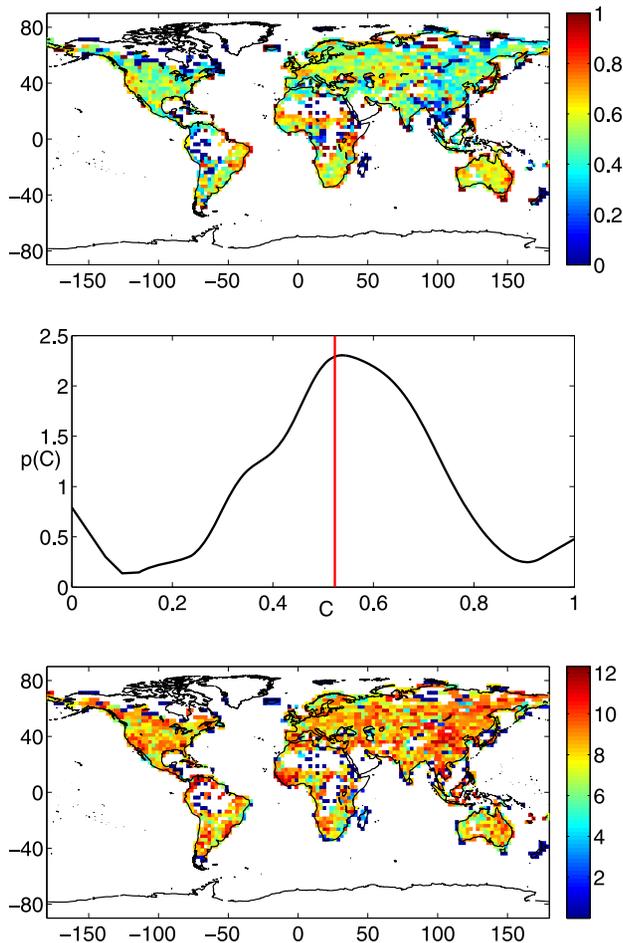

**Figure 6. Local clustering and betweenness centrality measures.** (top): local clustering coefficient, $C_i$. (middle): Probability density function, $p(C)$, of the local clustering coefficient, C. The red line represents the global clustering coefficient, $\bar{C} = 0.5233$. (bottom): betweenness centrality ($\ln(BC_i + 1)$ plot is shown).
doi:10.1371/journal.pone.0071129.g006

The betweenness centrality unveils the importance of a node in the network. Bottom panel of Fig. 6 summarizes that the nodes of the micro-networks and, more in general, nodes which are poorly connected to the rest of the network are the least important ones. These regions are depicted in dark blue. This result is not trivial, since the contrary (highly connected nodes are important) is not true. In fact, the importance of supernode areas and regions with low $\bar{D}_i$ values is not detectable at all from the betweenness centrality map.

The shortest path distribution for 4 significant nodes can be observed in Fig. 7. We recall that the shortest path length, $d_{ij}$, is the minimum number of edges that have to be crossed from node $i$ to node $j$. The four nodes are chosen as examples of relevant behaviors. Two nodes in the Northern and Southern Europe regions are shown in panels A and B, respectively. The Northern Europe node is closely linked to the whole European region and Western Russia. Meanwhile, a topological connection of the same strength is found with Northern and Central America. This pattern can be qualitatively associated to the Gulf Stream impact and to the atmospheric circulation induced by the North Atlantic Oscillation (NAO) [52]. Important connections are also visible with the Australian and African Sahel regions, while the farthest nodes are located in Eastern Asia. Although not physically distant from the Northern Europe node, the Southern Europe node presents a quite different scenario (see panel B). A strong connection is evident with the European and African Sahel regions only, while the links with the American and Australian continents are weaker. The node in the Southern part of America (panel C) is not deeply related to the confining Brazil region, but rather with the Caribbean America as well as with the Indonesian archipelago and Australia. Regions which are fairly linked with this node are the African Sahel, Western Russia and the Mongolian area. As a last example of shortest path length, we consider a node in Eastern Asia (panel D), which is a region with the highest weighted average topological distance (see the top panel of Fig. 4). The Eastern Asia node is topologically well connected only to those nodes which are also physically close to it. The whole European and American continents, which are physically distant, are furthermore vaguely linked to this node. In this case only, the topological distance is somehow related to the physical distance.

Some of the patterns linking mid-latitude to tropical nodes (e.g., North-American and European nodes with Sahel, Indonesian and Central America nodes) seem to suggest the impact of the propagation of planetary waves on precipitation at a global scale level [53]. Beside the influence of stationary planetary waves on the precipitation at local scale [54,55], it was recently found that extreme events simultaneously occur worldwide in concomitance with the amplification of trapped planetary waves [56].

Nevertheless, some links can be qualitatively related to the oceanic and atmospheric circulation [51]. In addition to the Gulf Stream and the NAO effects revealed by the shortest path of Northern Europe node (Fig. 7A), the South-Equatorial Current (linking the Pacific Coast of South America to Eastern Australia and Indonesian Archipelago) and the Brazil Current (linking the Atlantic Coast of South America to the Atlantic Coast of Africa) can be individuated for the South America node (Fig. 7C). The Australia node (see Fig. S1) is affected as well by the South-Equatorial Current branch going from Australia to South Africa. The Africa node (see Fig. S2) is related to the Atlantic Coast of South America through the Brazil Current.

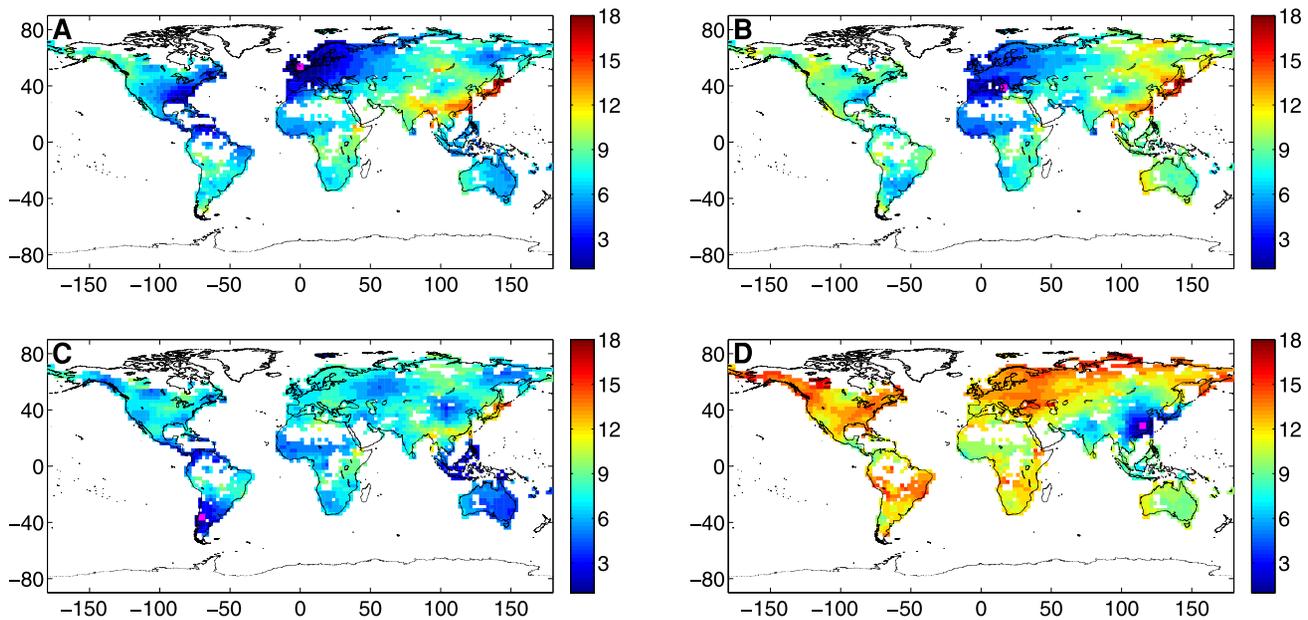

**Figure 7. Shortest path, $d_{ij}$, of 4 significant nodes of the network.** The measured nodes are represented in each panel by a pink square. (A) Northern Europe, node coordinates: (0° E, 53.75° N). (B) Southern Europe, node coordinates: (16.25° E, 38.75° N). (C) South America, node coordinates: (70° W, 36.25° S). (D) Asia, node coordinates: (115° E, 28.75° N).
doi:10.1371/journal.pone.0071129.g007

The different patterns expressed by the four nodes in Fig. 7 can be also observed through the physical area connected to each node as a function of the topological distance, see Fig. 8. Although the European trends are similar, the Southern node has significantly lower values in the range $d_{ij} \in [5,10]$. Moreover, there is a striking difference between the South America and Asia nodes. In fact, for $d_{ij} \in [4,9]$, the area connected to the South-American node is up to six/seven times larger than the area linked to the Asia node.

In (Text S1), an animated representation of the shortest path - linking a node to the rest of the network - is displayed for the nodes presented in Fig. 7 (see Movies S4, S5, S6, S7) and for other meaningful nodes of the network (see Figures S1, S2, S3 and Movies S1, S2, S3).

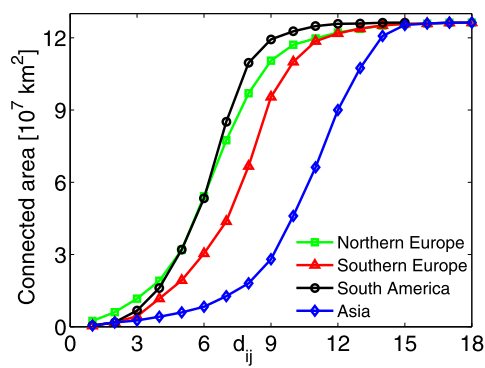

**Figure 8. Physical area connected to a node as a function of the topological distance, $d_{ij}$.** The four nodes of Fig. 7 are displayed.
doi:10.1371/journal.pone.0071129.g008

We conclude this section offering a possible climatological interpretation of two measures, the weighted average topological distance, $\bar{D}_i$, and the betweenness centrality, $BC_k$, which both rely on the concept of shortest path. As in complex network theory these two parameters are used to measure the information flow, here they should be intended as indexes of short- and long-range connections. Nodes with low $\bar{D}_i$ values are in general connected on a larger topological scale where precipitation varies more uniformly (e.g., Sahel region, Eastern Australia, Western Europe), while high $\bar{D}_i$ values describe regions with short-range connections, due to their higher precipitation variability (e.g., Southeastern Asia). The interpretation of the betweenness centrality is not so straightforward since it represents a mediator of both long- and short-range connections, which equally contribute to the final value of a node they pass through. As a consequence, the information on short- and long-range connections is partially lost. This is the reason why the betweenness centrality map is spotted, without remarkably patterned regions, and it is not very meaningful for the precipitation network.

## Conclusions

The recent development of complex network theory is offering new quantitative tools to disentangle the global climate dynamics. In spite of teleconnections having long been studied in climatology, the idea to read climatic correlations among different Earth regions as forming a complex network is relatively new. Starting from this point of view, we have focused on the global precipitation network. To this aim, we have used reliable datasets of measured land-surface precipitation that have only recently become available. Paying attention not to introduce spurious correlations due to uneven partitions of the Earth surface and to have a sufficient number of measured data in each cell, correlation analysis (with cut-off at 0.5) performed on a 70 year-window has yielded an undirected and symmetric network with 1674 nodes and 9481 links. We have investigated the structure of this precipitation network by some topological properties of nodes (degree centrality, local clustering coefficient, etc.) and, in particular, we have introduced a weighted form of the average topological distance in order to prevent some misleading behaviors when the graph has disconnected components.

Some key aspects of the precipitation network clearly emerge. Firstly, supernodes (i.e., highly connected nodes) occur in the Sahel region in Africa and in Eastern Australia, and a scaling-law behavior is revealed in the node degree distribution. Sahel and Eastern Australia regions are some of the most arid areas in the world. Very low rainfall is uniformly distributed on continental scales and huge extreme events are rare. As a consequence, the precipitation gradient tends to weaken, making these regions well connected on a large spatial scale. This long-range connection is confirmed by the fact that Sahel and Eastern Australia remain supernode regions also in the long-range network (see Fig. 5).

Strongly connected zones are evident also in Northern Europe, Central Asia, Western US, and Northeastern Brazil, even if they are not always characterized by a high betweenness centrality. Remarkably, strong connection differences occur between Northern and Mediterranean Europe as well as between Western and Eastern US. The European differences can be explained by the fact that Northern regions are mostly influenced by the oceanic circulation of the Gulf Stream and the atmospheric effects of the NAO (as observed in Fig. 7A), while Southern Europe is more affected by the Mediterranean, Saharan and Caucasian circulation (as shown in Fig. 7B). Differences for the US connections reflect the emergence of high precipitation gradients, which are partially due to frequent extreme events, and the ENSO influence on Western [32,57] and Eastern [58] US precipitation. Nevertheless, Eastern and Western Coasts are reached by very different ocean patterns: the Gulf Stream on the Atlantic Coast and the North Pacific Current on the Western Coast.

A second key point is the high topological distance of the Pacific region of the Asian continent with respect to the rest of the network. Notice that this behavior is not spuriously due to a geographic reason - namely, the decrease of lands above sea level close to the considered region - but it seems a peculiar feature of the precipitation network. A possible explanation of this network characteristic is the dramatic emergence of monsoons, tropical cyclones and heat waves which reduce the correlation only to the short-range scale. A similar disconnected behavior, even if less marked, is also observable for some coastal regions on the Southern Mediterranean and in Ethiopia. It should be noted as well that islands such as Iceland, Madagascar, New Zealand, and Japan, are completely disconnected from the rest of the network, forming different micro-networks. In fact, islands, which are in general more subject to extreme events [59], are not reached by the influence of the continental land mass and are furthermore exposed to the specific oceanic currents.

Finally, the shortest path distribution proved to be a powerful tool to unveil how information about precipitation in a node is linked to the network. Some geographic regions are embedded in very connected portions of the network and a few jumps between neighboring nodes are sufficient to cover large geographic regions; *vice versa*, other nodes turn out to be quite isolated from the rest of the network. The same tool is also useful to highlight the geography of this information flux about precipitation. For example, we have shown the case of two nodes in Europe that, in spite of their closeness, exhibit two rather different connection areas.

The reported results highlight that the complex network approach can be an useful framework to explore the huge amount of climatic data that have been collected in the last years and, then, shed light on climate dynamics.

## Supporting Information

**Text S1  Shortest path of significant nodes of the network.**
(PDF)

**Figure S1  Shortest path of the Australia node (coordinates: 141.25° E, 26.25° S) represented by a pink square.**
(EPS)

**Figure S2  Shortest path of the Africa node (coordinates: 3.75° W, 13.75° N) represented by a pink square.**
(EPS)

**Figure S3  Shortest path of the US node (coordinates: 105° W, 41.25° N) represented by a pink square.**
(EPS)

**Movie S1  Animated representation of the shortest path of the Australia node (coordinates: 141.25° E, 26.25° S) displayed in Fig. S1.** The node is indicated by a pink square.
(AVI)

**Movie S2  Animated representation of the shortest path of the Africa node (coordinates: 3.75° W, 13.75° N) displayed in Fig. S2.** The node is indicated by a pink square.
(AVI)

**Movie S3  Animated representation of the shortest path of the US node (coordinates: 105° W, 41.25° N) displayed in Fig. S3.** The node is indicated by a pink square.
(AVI)

**Movie S4  Animated representation of the shortest path of the Northern Europe node (coordinates: 0° E, 53.75° N) displayed in Fig. 7A of the main text.** The node is indicated by a pink square.
(AVI)

**Movie S5  Animated representation of the shortest path of the Southern Europe node (coordinates: 16.25° E, 38.75° N) displayed in Fig. 7B of the main text.** The node is indicated by a pink square.
(AVI)

**Movie S6  Animated representation of the shortest path of the South America node (coordinates: 70° W, 36.25° S) displayed in Fig. 7C of the main text.** The node is indicated by a pink square.
(AVI)

**Movie S7  Animated representation of the shortest path of the Asia node (coordinates: 115° E, 28.75° N) displayed in Fig. 7D of the main text.** The node is indicated by a pink square.
(AVI)

## Author Contributions

# References

1. Watts DJ, Strogatz SH (1998) Collective dynamics of 'small-world' networks. Nature 393: 440442. (DOI 10.1038/30918.).

2. Newman MEJ (2003) The structure and function of complex networks. SIAM Rev 45: 167256. (DOI 10.1137/S003614450342480.).

3. Albert R, Barabási AL (2002) Statistical mechanics of complex networks. Rev Mod Phys 74 (1): 47–97 (DOI 10.1103/RevModPhys.74.47.).

4. Costa LD, Oliveira ON, Travieso G, Rodrigues FA, Boas PRV et al. (2011) Analyzing and modeling real-world phenomena with complex networks: a survey of applications. Adv Phys 60 (3): 329–412 (DOI 10.1080/00018732.2011.572452.).

5. Boccaletti S, Latora V, Moreno Y, Chavez M, Hwang DU (2006) Complex networks: Structure and dynamics. Phys Rep 424: 175–308 (DOI 10.1016/j.physrep.2005.10.009.).

6. Davis KF, D'Odorico P, Laio F, Ridolfi L (2013) Global Spatio-Temporal Patterns in Human Migration: A Complex Network Perspective. PLOS ONE 8 (1): e53723 (DOI 10.1371/journal. pone.0053723.).

7. Chen LL, Blumm N, Christakis NA, Barabási AL, Deisboeck TS (2009) Cancer metastasis networks and the prediction of progression patterns. Brit J Cancer 101: 749758 (DOI 10.1038/sj.bjc.6605214.).

8. Abe S, Suzuki N (2004) Scale-free network of earthquakes. Europhys Lett 65 (4): 581586 (DOI 10.1209/epl/i2003-10108-1.).

9. Palus M, Hartman D, Hlinka J, Vejmelka M (2011) Discerning connectivity from dynamics in climate networks. Nonlin Processes Geophys 18: 751763 (DOI 10.5194/npg-18-751-2011.).

10. Donges JF, Zou Y, Marwan N, Kurths J (2009) Complex networks in climate dynamics. Eur Phys J Special Topics 174: 157–179 (DOI 10.1140/epjst/e2009-01098-2.).

11. Donges JF, Zou Y, Marwan N, Kurths J (2009) The backbone of the climate network. Europhys Lett 87: 48007 (DOI 10.1209/0295-5075/87/48007.).

12. Yamasaki K, Gozolchiani A, Havlin S (2008) Climate networks around the globe are significantly affected by El Niño. Phys Rev Lett 100: 228501 (DOI 10.1103/PhysRevLett.100.228501.).

13. Tsonis AA, Swanson KL (2008) Topology and Predictability of El Niño and La Niña Networks. Phys Rev Lett 100: 228502 (DOI 10.1103/PhysRevLett.100.228502.).

14. Gozolchiani A, Yamasaki K, Gazit O, Havlin S (2008) Pattern of climate network blinking links follows El Niño events. Europhys Lett 83: 28005 (DOI 10.1209/0295-5075/83/28005.).

15. Tsonis AA, Wang G, Swanson KL, Rodrigues FA, Costa LD (2011) Community structure and dynamics in climate networks. Clim Dyn 37: 933940 (DOI 10.1007/s00382-010-0874-3.).

16. Donges JF, Schultz HCH, Zou Y, Marwan N, Kurths J (2011) Investigating the topology of interacting networks. Eur Phys J Special Topics 84: 635651 (DOI 10.1140/epjb/e2011-10795-9.).

17. Tsonis AA, Swanson KL (2008) On the Role of Atmospheric Teleconnections in Climate. J Climate 21: 2990–3001 (DOI 10.1175/2007JCLI1907.1.).

18. Steinhaeuser K, Ganguly AR, Chawla NV (2012) Multivariate and multiscale dependence in the global climate system revealed through complex networks. Clim Dyn 39: 889895 (DOI 10.1007/s00382-011-1135-9.).

19. Malik N, Bookhagen B, Marwan N, Kurths J (2012) Analysis of spatial and temporal extreme monsoonal rainfall over South Asia using complex networks. Clim Dyn 39: 971987 (DOI 10.1007/s00382-011-1156-4.).

20. Cano R, Sordo C, Gutiérrez JM (2004) Applications of Bayesian Networks in Meteorology. Advances in Bayesian Networks 146: 309–327.

21. Cofiño AS, Cano R, Sordo C, Gutiérrez JM (2002) Bayesian Networks for Probabilistic Weather Prediction. Proceedings of the 15th European Conference on Artificial Intelligence (ECAI 2002) 77: 695–699.

22. van Oldenborgh GJ, Burgers G (2005) Searching for decadal variations in ENSO precipitation teleconnections. Geophys Res Lett 32 (15): L15701 (DOI 10.1029/2005GL023110.).

23. Diaz JF, Hoerling MP, Eischeid JK (2001) ENSO variability, teleconnections and climate change. Int J Climatol 21: 1845 1862 (DOI 10.1002/joc.631.).

24. Archer DR, Fowler HJ (2004) Spatial and temporal variations in precipitation in the Upper Indus Basin, global teleconnections and hydrological implications. Hydrol Earth Syst Sc 8 (1): 47–61 (DOI 10.5194/hess-8-47-2004.).

25. Zhou T, Zhang L, Li H (2008) Changes in global land monsoon area and total rainfall accumulation over the last half century. Geophys Res Lett 35: L16707 (DOI 10.1029/2008GL034881.).

26. Saeed S, Müller WA, Hagemann S, Jacob D, Mujumdar M et al. (2011) Precipitation variability over the South Asian monsoon heat low and associated teleconnections. Geophys Res Lett 38: L08702 (DOI 10.1029/2011GL046984.).

27. Raicich F, Pinardi N, Navarra A (2003) Teleconnections between Indian Monsoon and Sahel rainfall and the Mediterranean. Int J Climatol 23: 173186 (DOI 10.1002/joc.862.).

28. Kripalani RH, Kulkarni A (2001) Monsoon rainfall variations and teleconnections over South and East Asia. Int J Climatol 21: 603616 (DOI 10.1002/joc.625.).

29. Shaman J, Tziperman E (2011) An Atmospheric Teleconnection Linking ENSO and Southwestern European Precipitation. J Climate 24: 124–139 (DOI 10.1175/2010JCLI3590.1.).

30. Brönnimann S (2007) Impact of El NiñoSouthern Oscillation on European climate. Rev Geophys 45: RG3003 (DOI 10.1029/2006RG000199.).

31. Nicholson SE, Kim E (1997) The relationship of the El Nino Southern oscillation to African rainfall. Int J Climatol 17 (2): 117–135 (DOI 10.1002/(SICI)1097-0088(199702)17:2<117::AIDJOC43>3.0.CO;2-O.).

32. McCabe GJ, Dettinger MD (1999) Decadal variations in the strength of Enso teleconnections with precipitation in the Western United States. Int J Climatol 19 (13): 13991410 (DOI 10.1002/(SICI)1097-0088(19991115)19:13<1399::AID-JOC457>3.0.CO;2-A.).

33. Boulanger JP, Leloup J, Penalba O, Rusticucci M, Lafon F et al. (2005) Observed precipitation in the Paraná-Plata hydrological basin: long-term trends, extreme conditions and ENSO teleconnections. Clim Dyn 24: 393413 (DOI 10.1007/s00382-004-0514-x.).

34. Schneider U, Becker A, Finger P, Meyer-Christoffer A, Ziese M et al. (2013) GPCC's new land surface precipitation climatology based on quality-controlled in situ data and its role in quantifying the global water cycle. Theor Appl Climatol (DOI 10.1007/s00704-013-0860-x.).

35. Becker A, Finger P, Meyer-Christoffer A, Rudolf B, Schamm K et al. (2013) A description of the global land-surface precipitation data products of the Global Precipitation Climatology Centre with sample applications including centennial (trend) analysis from 1901present. Earth Syst Sci Data 5: 71–99 (DOI 10.5194/essd-5-71-2013.).

36. GPCC Homepage: http://gpcc.dwd.de.

37. Hemming D, Buontempo C, Burke E, Collins M, Kaye N (2010) How uncertain are climate model projections of water availability indicators across the Middle East? Philos T Roy Soc A 368 (1931): 5117–5135 (DOI 10.1098/rsta.2010.0174.).

38. Fritsche M, Döll P, Dietrich R (2012) Global-scale validation of model-based load deformation of the Earth's crust from continental watermass and atmospheric pressure variations using GPS. J Geodyn 59-60: 133142 (DOI 10.1016/j.jog.2011.04.001.).

39. Michaelides S, Levizzani V, Anagnostou E, Bauer P, Kasparis T et al. (2009) Precipitation: Measurement, remote sensing, climatology and modeling. Atmos Res 94 (4): 512533 (DOI 10.1016/j.atmosres.2009.08.017.).

40. Kennedy J, Titchner H, Hardwick J, Beswick M, Parker DE (2007) Global and regional climate in 2006. Weather 62: 232242 (DOI 10.1002/wea.119.).

41. Kennedy J, Titchner H, Parker D, Beswick M, Hardwick J et al. (2008) Global and regional climate in 2007. Weather 63: 296304 (DOI 10.1002/wea.320.).

42. Kennedy J, Morice C, Parker D (2012) Global and regional climate in 2011.Weather 67 8: 212–218 (DOI 10.1002/wea.1945.).

43. Fatichi S, Ivanov VY, Caporali E (2012) Investigating Interannual Variability of Precipitation at the Global Scale: Is There a Connection with Seasonality? J Climate 25 (16): 5512–5523 (DOI 10.1175/JCLI-D-11-00356.1.).

44. Chou C, Chiang JC, Lan CW, Chung CH, Liao YC et al. (2013) Increase in the range between wet and dry season precipitation. Nat Geosci 6: 263267 (DOI 10.1038/ngeo1744.).

45. Gessner U, Naeimi V, Klein I, Kuenzer C, Klein D et al. (2012) The relationship between precipitation anomalies and satellite-derived vegetation activity in Central Asia. Global Planet Change (DOI 10.1016/j.gloplacha.2012.09.007.).

46. Simmons AJ, Willett KM, Jones PD, Thorne PW, Dee DP (2010) Low-frequency variations in surface atmospheric humidity, temperature, and precipitation: Inferences from reanalyses and monthly gridded observational data sets. J Geophys Res Atmos 115 (1): D01110 (DOI 10.1029/2009JD012442.).

47. Kolzsch A, Saether SA, Gustafsson H, Fiske P, Hoglund J et al. (2007) Population fluctuations and regulation in great snipe: a time-series analysis. J Anim Ecol 76 (4): 740–749 (DOI 10.1111/j.1365-2656.2007.01246.x.).

48. Costa LD, Rodrigues FA, Travieso G, Boas PRV (2007) Characterization of complex networks: A survey of measurements. Adv Phys 56 (1): 167–242 (DOI 10.1080/00018730601170527.).

49. Heitzig J, Donges JF, Zou Y, Marwan N, Kurths J (2012) Node-weighted measures for complex networks with spatially embedded, sampled, or differently sized nodes. Eur Phys J B 85: 38 (DOI 10.1140/epjb/e2011-20678-7.).

50. Freeman LC (1979) Centrality in social networks conceptual clarification. Soc Networks 1 (3): 215239 (DOI 10.1016/0378-8733(78)90021-7.).

51. Trewartha GT, Horn LH (1980) An Introduction to Climate. McGraw-Hill, New York-London.

52. Hurrell JW (1995) Decadal Trends in the North Atlantic Oscillation: Regional Temperatures and Precipitation. Science 269 (5224): 676–679 (DOI 10.1126/science.269.5224.676.).

53. Hansen JE, Takahashi T (1984) Climate Processes and Climate Sensitivity. Geophys Monogr Ser 29: 368 AGU, Washington, DC. (DOI 10.1029/GM029.).

54. Chen W, Yang S, Huang RH (2005) Relationship between stationary planetary wave activity and the East Asian winter monsoon. J Geophys Res 110: D14110 (DOI 10.1029/2004JD005669.).

55. Ryoo JM, Kaspi Y, Waugh DW, Kiladis GN, Waliser ED et al. (2013) Impact of Rossby wave breaking on U.S. west coast winter precipitation during ENSO events. J Climate (DOI 10.1175/JCLID-12-00297.1.).

56. Petoukhova V, Rahmstorfa S, Petria S, Schellnhuber HJ (2013) Quasiresonant amplification of planetary waves and recent Northern Hemisphere weather

extremes. P Natl Acad Sci Usa 110 (14): 5336–5341 (DOI 10.1073/pnas.1222000110.).

57. Rajagopalan B, Lall U (1998) Interannual variability in western US precipitation. J Hydrol 210 (1–4): 51–67 (DOI 10.1016/S0022-1694(98)00184-X.).

58. Montroy DL, Richman MB, Lamb PJ (1998) Observed nonlinearities of monthly teleconnections between tropical Pacific sea surface temperature anomalies and central and eastern North American precipitation. J Climate 11 (7): 1812–1835 (DOI: 10.1175/1520–0442(1998)011<1812:ONOMTB>2.0.CO;2.).

59. Barros V, Field CB, Dahe Q, Stocker TF (2012) Managing the Risks of Extreme Events and Disasters to Advance Climate Change Adaptation. Cambridge University Press, New York.